\documentclass[a4paper, 12pt]{article}\usepackage[]{graphicx}\usepackage[]{color}
\makeatletter
\def\maxwidth{ %
  \ifdim\Gin@nat@width>\linewidth
    \linewidth
  \else
    \Gin@nat@width
  \fi
}
\makeatother

\definecolor{fgcolor}{rgb}{0.345, 0.345, 0.345}
\makeatletter
\@ifundefined{AddToHook}{}{\AddToHook{package/xcolor/after}{\definecolor{fgcolor}{rgb}{0.345, 0.345, 0.345}}}
\makeatother

\usepackage{framed}
\makeatletter
 {\par\unskip\endMakeFramed%
 \at@end@of@kframe}
\makeatother

\definecolor{shadecolor}{rgb}{.97, .97, .97}
\definecolor{messagecolor}{rgb}{0, 0, 0}
\definecolor{warningcolor}{rgb}{1, 0, 1}
\definecolor{errorcolor}{rgb}{1, 0, 0}
\makeatletter
\@ifundefined{AddToHook}{}{\AddToHook{package/xcolor/after}{
\definecolor{shadecolor}{rgb}{.97, .97, .97}
\definecolor{messagecolor}{rgb}{0, 0, 0}
\definecolor{warningcolor}{rgb}{1, 0, 1}
\definecolor{errorcolor}{rgb}{1, 0, 0}
}}
\makeatother
\newenvironment{knitrout}{}{} 

\usepackage{alltt}
\usepackage{amsmath, amssymb, amsthm}
\usepackage[numbers, sort&compress]{natbib}
\usepackage[english]{babel}
\usepackage[utf8]{inputenc}
\usepackage{url}
\usepackage{float}
\usepackage{graphicx}
\usepackage[pdftex, pdfpagelabels]{hyperref}
\usepackage{geometry}
\usepackage{dsfont} 
\usepackage{todonotes}

\theoremstyle{definition}

\newtheorem{remark}{Remark}[section]

\newcommand{\monthname}{\ifcase\month \or January\or February\or March\or 
April\or May \or June\or July\or August\or September\or October\or November\or 
December\fi}
\newcommand{\R}{\texttt{R }} 
\newcommand{\latin}[1]{\textit{#1}} 
\newcommand{\abk}[1]{\mbox{#1}\xdot} 
\DeclareRobustCommand\xdot{\futurelet\token\Xdot} 
\def\Xdot{%
  \ifx\token\bgroup.%
  \else\ifx\token\egroup.%
  \else\ifx\token\/.%
  \else\ifx\token\ .%
  \else\ifx\token!.%
  \else\ifx\token,.%
  \else\ifx\token:.%
  \else\ifx\token;.%
  \else\ifx\token?.%
  \else\ifx\token/.%
  \else\ifx\token'.%
  \else\ifx\token).%
  \else\ifx\token-.%
  \else\ifx\token+.%
  \else\ifx\token~.%
  \else\ifx\token.%
  \else.\ %
  \fi\fi\fi\fi\fi\fi\fi\fi\fi\fi\fi\fi\fi\fi\fi\fi%
}
\newcommand{\eg}{\abk{\latin{e.\,g}}}

\newcommand{\pihat}{\widehat{\pi}_{\scriptscriptstyle{\mathrm{ML}}}}
\newcommand{\phihat}{\widehat{\phi}_{\scriptscriptstyle{\mathrm{ML}}}}
\newcommand{\sima}{\mathrel{\overset{a}{\thicksim}}} 
\newcommand{\given}{\,\vert\,} 
\DeclareMathOperator{\N}{N} 
\DeclareMathOperator{\Bin}{Bin} 
\DeclareMathOperator{\Unif}{U} 
\DeclareMathOperator{\Be}{Be} 
\DeclareMathOperator{\Prob}{\mathsf{Pr}} 
\DeclareMathOperator{\Exp}{\mathsf{E}} 
\DeclareMathOperator{\se}{se} 
\DeclareMathOperator{\logit}{logit}
\DeclareMathOperator{\expit}{expit}
\DeclareMathOperator{\ind}{\mathds{1}} 
\DeclareMathOperator{\IS}{IS}

\DeclareMathOperator{\EIS}{EIS}

\DeclareMathOperator{\CP}{CP}
\DeclareMathOperator{\C}{C}
\DeclareMathOperator{\EW}{EW}
\DeclareMathOperator{\W}{W}

\title{\textbf{Comparing Confidence Intervals for a Binomial Proportion with the Interval Score}}
\author{Lisa J. Hofer \and Leonhard Held}
\date{Epidemiology, Biostatistics and Prevention Institute (EBPI),
      University of Zurich, Switzerland\\
      Email: \href{mailto:lisa.hofer@uzh.ch}{lisa.hofer@uzh.ch}\\[2ex]
      \today}
\IfFileExists{upquote.sty}{\usepackage{upquote}}{}
\begin{document}
\maketitle


\begin{abstract}
There are over $55$ different ways to construct a confidence respectively credible interval (CI) for the binomial proportion. Methods to compare them are necessary to decide which should be used in practice. The interval score has been suggested to compare prediction intervals. This score is a proper scoring rule that combines the coverage as a measure of calibration and the width as a measure of sharpness. We evaluate eleven CIs for the binomial proportion based on the expected interval score and propose a summary measure which can take into account different weighting of the underlying true proportion. Under uniform weighting, the expected interval score recommends the Wilson CI or Bayesian credible intervals with a uniform prior. If extremely low or high proportions receive more weight, the score recommends Bayesian credible intervals based on Jeffreys' prior. While more work is needed to theoretically justify the use of the interval score for the comparison of CIs, our results suggest that it constitutes a useful method to combine coverage and width in one measure. This novel approach could also be used in other applications.\newline

\end{abstract}

\section{Introduction} \label{sec:intro}

It is good scientific practice to report the point estimate together with a confidence interval (CI) \citep{Altman}. The advantage of this interval estimate over the point estimate or a $p$-value is that it provides information about the magnitude and the precision of the estimate at once \citep{Rothman}. However, especially for estimating binomial proportions which is one of the oldest statistical problems, it is difficult to construct CIs with good properties. Because the binomial distribution is bounded and discrete, many different interval estimators have been developed using different ways of approximation. The question then is which to choose. Consequently, there is a vast literature on comparisons of CI methods for the binomial proportion. These comparisons differ both in the selection of estimators as well as in the choice of evaluation criteria.

For illustration, we consider a recent example on the reliability of COVID-19 self-tests. This is measured by the ability of the test to detect an infection when it is present (sensitivity) but also to correctly identify when there is no infection (specificity). These two quantities are the binomial proportions of correct test results among participants with (sensitivity) or without (specificity) an infection where the correctness is assessed against a PCR test considered the gold standard. \citet{Lindner} reported a sensitivity of $74.4\%$ ($95\%$ CI $58.9$ to $85.4$) and a specificity of $99.2\%$ ($95\%$ CI $97.1$ to $99.8$) based on data summarized in Table~\ref{example}.
\begin{table}[h]
\caption{Results of a diagnostic study about COVID-19 self-tests with different $95\%$ confidence intervals for sensitivity and specificity from \cite{Lindner}.}
\label{example}
\begin{center}
\begin{tabular}{l r}
\begin{tabular}{l|rr|r}
 & Covid & & \\
Self-test  & no & yes & Total   \\
\hline
negative & $246$ & $10$ & $256$ \\
positive & $2$ & $29$ & $31$ \\
\hline
Total & $248$ & $39$ & $287$
\end{tabular}
\hfill &
\begin{tabular}{l|l|l}
CI & & \\
method  & Sensitivity & Specificity   \\
\hline
Wald & 60.7 to 88.1 &  98.1 to 100.3 \\
Wilson & 58.9 to 85.4 &  97.1 to  99.8 \\
Clopper-Pearson & 57.9 to 87.0 &  97.1 to  99.9
\end{tabular}
\end{tabular}
\end{center}
\end{table}
Table~\ref{example} also shows three of the most commonly used CI methods in this example. The study reported the Wilson CI. Unlike the Wald CI, it avoids overshoot below $0\%$ or above $100\%$ as here for the specificity. Another boundary anomaly of the Wald CI is a zero-width interval that occurs whenever there are no events or non-events. In this example, the limits of the Wald CI for sensitivity are around $2$ percentage points larger than those of the Wilson CI. The Clopper-Pearson CI on the other hand is more conservative as it is wider and contains the Wilson CI.

For the last 30 years, researchers have criticized that the Wald CI is used for binomial proportions although it is known for boundary anomalies and coverage bias. \citet{Vollset} compared coverage probabilities and expected widths as a function of the true proportion with fixed sample size. Others looked at mean or minimum coverage probability for fixed sample size in order to summarize over all possible true proportions \citep{Newcombe1998,Brown2001,PiresAmado}. Sometimes also interval location is a criterion in terms of similar left and right non-coverage \citep{Newcombe2013,Gillibert}. As a systematic review, the \citet{Gillibert} comparison is the largest with 55 different CIs. CIs that have often been recommended are the Wilson, Agresti-Coull or Bayesian equal-tailed with a uniform or Jeffreys' prior. The recommendations depend on what evaluation criteria are used, but most commonly coverage and width is the focus.

A major issue is that coverage and width are assessed separately but there is a trade-off. The best interval would be as small as possible while still respecting the correct coverage. This is a trade-off because decreasing the interval width decreases the coverage and increasing the coverage increases the interval width. This relation is referred to as \emph{sharpness subject to calibration}, where the coverage calibrates the interval while the width determines its sharpness \citep{GneitingBalabdaoui}. A way to assess calibration and sharpness simultaneously are scoring rules. Such a scoring rule is the interval score that combines coverage and width in a loss function. This score has been developed for prediction intervals but the authors suggest that it could also be used for interval estimates, intended to compare intervals for the same nominal coverage that have equal lower and upper exceedance probabilities which they call \emph{central} \citep[p.~18]{GneitingRaftery}. Only in this case, the interval score is a \emph{proper} scoring rule such that the optimal interval estimate minimizes the expected score. While the idea of a method for a combined assessment of coverage and width for CIs existed, it has not yet been applied. In particular, it is not straightforward what the concept of proper scoring rules would mean in this new setting.

Proper scoring rules are usually employed to compare probabilistic predictions, \eg quantile predictions \citep{GneitingRaftery}. In this context, a scoring rule is proper if the expected score is optimized under, \eg, the true quantiles. On the other hand, they are also used to calculate point estimates, \eg the log score in maximum likelihood estimation or the continuous ranked probability score \citep{Gneiting2005,Hothorn}. In this context, a loss function is a proper score if the expected loss is optimized at the true parameter point \citep{GneitingRaftery,Buja}. The interval score is a proper scoring rule for \emph{central} prediction intervals. Using the interval score to compare interval estimates is yet another application to which the known definitions of propriety cannot be easily translated. Binomial proportions with a discrete outcome, \eg, are a case where an optimal CI does not exist but it would be needed to check for propriety. Another difference is that the frequentist setting does not assume a probability distribution for the unknown parameter. Nonetheless, borrowing proper scoring rules from a prediction setting has the advantage that it avoids paradoxes, \eg a paradox from \cite{CasellaHwangRobert} discussed in \cite[p.~18]{GneitingRaftery}.

In this article, we address the trade-off between coverage and width by using the expected interval score as a new measure to compare CIs. We apply it in the example of binomial proportions for a selection of eleven commonly used CIs: Clopper-Pearson, Wilson, Wald, Rindskopf, arcsine Wald, Agresti-Coull, likelihood ratio and Bayesian equal-tailed and HPD intervals with uniform and Jeffreys priors. We use the integral of the expected interval score over all possible true proportions $\pi$ as a summary measure that corresponds to the mean coverage probability assuming a uniform distribution for $\pi$ \citep[p.~104--105]{Newcombe2013}. To attribute more weight to extreme cases, we also integrate with respect to the variance-stabilizing transformation of $\pi$. A good performance in extreme cases is important in particular for sensitivity and specificity as they are usually very high. Finally, a new asymptotical result on the expected interval score helps to visualize the results. Using this novel approach to evaluate CIs, we obtain a clear ranking.

This article extends previous work by \citet{HoferMA} and is structured as follows: Setting and notation as well as the used methods are described in Section~\ref{sec:methods}, followed by the results in Section~\ref{sec:res} and a discussion in Section~\ref{sec:discussion}. The appendix contains information about computation of the CIs.

\section{Methods} \label{sec:methods}

The parameter of interest is the unknown success probability $\pi\in (0,1)$ of a binomial sample $X\sim \Bin(n, \pi)$, where $X$ denotes the number of successes and $n$ is the known sample size. A realization of the random variable $X$ is denoted by $x$. Because $n\pi$ is the expected number of successes in the sample, $\pi$ is the expected proportion of successes often referred to as the \emph{binomial proportion}.

\subsection{Confidence and credible intervals for the binomial proportion}

We consider both frequentist and Bayesian interval estimators for the unknown proportion $\pi$ and evaluate their frequentist properties, where $\pi$ is assumed to be fixed. First, let $\gamma \in (0,1)$ be the confidence level and its complement be $\alpha = 1-\gamma$. A $\gamma\cdot 100\%$ confidence interval (CI) for $\pi$ is defined as an interval $[L,U]$ that fulfills
\begin{equation*}
\Prob(L\leq \pi \leq U) = \gamma.
\end{equation*}
Since $\pi$ is fixed, no probability statement is attached to $\pi$ but to the limits $L$ and $U$. For repeated random samples $X\sim \Bin(n, \pi)$, a $\gamma\cdot 100\%$ CI will cover the parameter $\pi$ in $\gamma\cdot 100\%$ of all cases. Realizations of the random variables $L$ and $U$ are denoted by $l$ and $u$.

In a Bayesian context, $\pi$ has a distribution. From an assumed prior distribution and an observation $x$, a posterior distribution with density $f(\pi \given x)$ is calculated. A credible interval $[l,u]$ for $\pi$ with credibility level $\gamma$ is defined by two quantiles $l$ and $u$ of the posterior distribution that fulfill
\begin{equation*}
\int_{l}^{u} f(\pi \given x) d\pi = \gamma.
\end{equation*}
Under the assumed prior distribution, the random variable $\pi \given x$ is contained in a ${\gamma\cdot 100\%}$ credible interval with probability $\gamma$. Although this is conceptually different than frequentist inference, we interpret credible intervals as confidence intervals (and abbreviate them also by CI) as it has been also done \eg in \cite{PiresAmado} and \cite{BayarriBerger}.

The CIs that we compare in this article are listed in Table~\ref{intervals}. 
\begin{table}[h]
\caption{Considered confidence and credible intervals for the binomial proportion.}
\label{intervals}
\begin{center}
\begin{tabular}{l|l}
 Frequentist CIs & Bayesian CIs \\
 \hline
 Wald & Jeffreys equal-tailed \\
 Rindskopf & Jeffreys HPD \\
 Arcsine Wald & Uniform equal-tailed \\
 Wilson & Uniform HPD \\ 
 Agresti-Coull &  \\
 Likelihood ratio &  \\
 Clopper-Pearson &  
\end{tabular}
\end{center}
\end{table}
Details about computation and motivation can be found in Appendix~\ref{sec:intervals}. We evaluate all interval estimators using frequentist properties such as the coverage probability. The perfect interval estimator has coverage probability $\gamma$ which is called the \emph{nominal} coverage probability. However, since the binomial distribution is discrete, all interval estimators will only approximately have the intended coverage probability. Bayesian intervals have exact mean coverage probability equal to $\gamma$ when integrated under the specified prior distribution \citep{PiresAmado}. Intervals using the Jeffreys or the uniform prior are known to have favourable frequentist properties \citep{BayarriBerger,Newcombe2013}. These priors are non-informative, hence conceptually as close to the frequentist setting as possible.

\begin{remark}
A $\gamma\cdot 100\%$ CI is \emph{central} if the lower and upper \emph{exceedance probabilities} (left and right \emph{non-coverage probabilities} in \cite{Newcombe1998}) are both $\alpha/2$ \citep[Subsection~9.3]{GneitingRaftery}. For non-central intervals, the interval score is not a proper scoring rule \citep{GneitingBrehmer}. All frequentist intervals in Table~\ref{intervals} are intended to be central and are interpreted as such. They are also considered as central in \cite{PiresAmado}. However, since the calculations rely on approximations, they are only asymptotically central. The Bayesian equal-tailed interval is central in a rigorous way. The HPD interval (considered as non-central in \cite{PiresAmado}) is only asymptotically central.
\end{remark}

\subsection{Interval evaluation methods}

\begin{figure}
\begin{knitrout}
\definecolor{shadecolor}{rgb}{0.969, 0.969, 0.969}\color{fgcolor}

{\centering \includegraphics[width=\maxwidth]{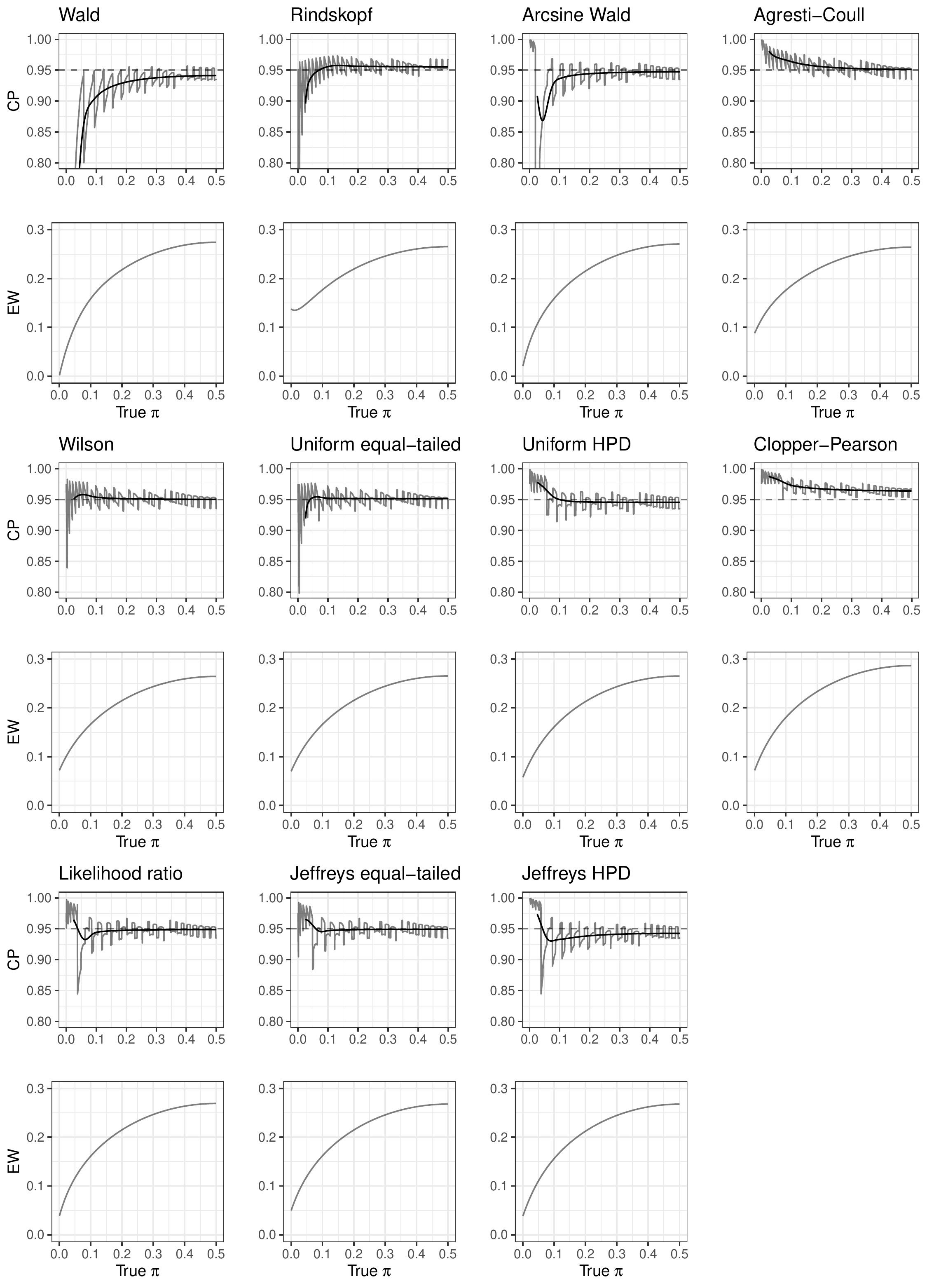} 

}

\end{knitrout}
\caption{Coverage probability (CP) and expected width (EW) for each confidence interval method as a function of $\pi$ for $n = 50$ and $\gamma = 0.95$. Locally smoothed coverage probability added in black.}
\label{covwidth}
\end{figure}

Since a binomial sample $X\sim \Bin(n, \pi)$ only has a finite number of possible outcomes, coverage probabilities and expected widths of a certain CI can be calculated analytically \citep{Vollset,PiresAmado}. As a function of the true proportion $\pi$, they are symmetric around $\pi=0.5$ due to the equivariance of lower and upper bounds of the CIs \citep{Gillibert}. Figure~\ref{covwidth} shows coverage probabilities and expected widths of the discussed CIs as a function of $\pi$ in a binomial experiment with $n=50$ and $\gamma=0.95$. We added smoothed coverage probabilities (in black) that are computed using a specific kernel function as described in \cite{BayarriBerger} for the smoothing parameter ${\varepsilon = 0.025}$. Smoothed coverage probabilities help to compare the CIs because the unsmoothed coverage probabilities oscillate due to the discreteness of the binomial distribution.

\subsubsection{Coverage probability}
The \emph{coverage} of a CI $[l,u]$ for a true proportion $\pi$ is
\begin{equation*}
\C(l, u, \pi) = \ind(l \leq \pi \leq u),
\end{equation*}
where $\ind$ denotes the indicator function. The \emph{coverage probability} is the probability with which a random CI contains the true proportion $\pi$:
\begin{align*}
\begin{split}
\CP(\pi) &= \sum_{x=0}^{n}\Prob(X=x)\C(l(x),u(x), \pi)\\
&= \sum_{x=0}^{n}{n \choose x} \pi^{x} (1-\pi)^{n-x} \C(l(x),u(x), \pi).
\end{split}
\end{align*}

\subsubsection{Expected width}
The \emph{width} of a CI $[l,u]$ for a true proportion $\pi$ is
\begin{equation*}
\W(l, u) = u-l
\end{equation*}
and the \emph{expected width} (expectation w.r.t.\ the distribution of the data $X$) is
\begin{equation*}
\EW(\pi) = \sum_{x=0}^{n}\Prob(X = x) \W(l(x),u(x)).
\end{equation*}

\subsubsection{Expected interval score}
The \emph{interval score} of a $\gamma\cdot 100\%$ CI $[l,u]$ for a true proportion $\pi$ is
\begin{align}\label{eq:IS}
\begin{split}
\IS_\alpha(l, u, \pi) 
&= (u-l) + \frac{2}{\alpha}(l-\pi)\ind(\pi < l) + \frac{2}{\alpha}(\pi-u)\ind(\pi > u)\\
&= \W(l, u) + \frac{2}{\alpha}\min(\lvert \pi-l \rvert, \lvert \pi-u \rvert) \Bigr\{1 - \C(l, u, \pi)\Bigr\},
\end{split}
\end{align}
where $\alpha=1-\gamma$. This score is introduced in \cite[Section~6]{GneitingRaftery} for central prediction intervals as a special case of a more general proper scoring rule for predictive quantiles. The second line of \eqref{eq:IS} expresses the interval score in terms of two penalties. One penalty is the interval width where a larger interval is a larger penalty. The other penalty is given for non-coverage. This penalty of value $1$ is weighted by the amount of non-coverage, more precisely by the minimal distance of the true proportion to the interval and by the reciprocal of half the exceedance probability. Width and coverage are combined into a negatively oriented score, meaning that lower scores are better. The \emph{expected interval score} of a $\gamma\cdot 100\%$ CI is
\begin{equation*}
\EIS_\alpha(\pi) = \sum_{x=0}^{n}\Prob(X = x)\IS_\alpha(l(x), u(x), \pi).
\end{equation*}

\subsection{Integral summary measures} \label{subsec:integral}

Coverage probability, expected width and expected interval score depend on the assumed true proportion $\pi$. The integral over all possible true proportions is a summary measure that can be compared across CIs for fixed $n$ and $\gamma$. It is closely related to the concept of \emph{integrated risk} in Bayesian decision theory \citep[p.~62--63]{Robert} which integrates the frequentist risk (\eg the expected interval score) over $\pi$ with respect to the prior distribution of $\pi$.

\begin{figure}
\begin{knitrout}
\definecolor{shadecolor}{rgb}{0.969, 0.969, 0.969}\color{fgcolor}

{\centering \includegraphics[width=\maxwidth]{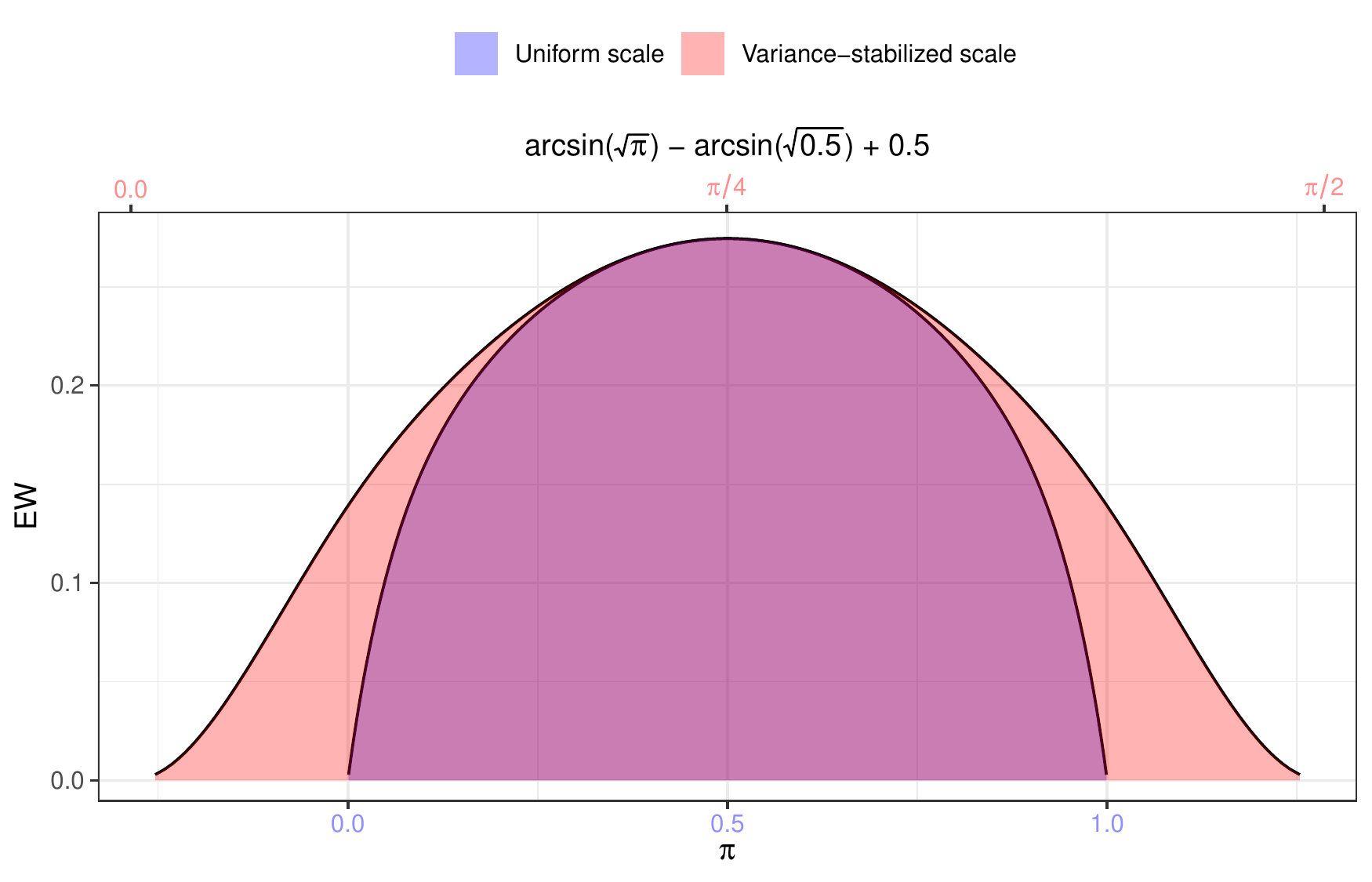} 

}

\end{knitrout}
\caption{Integral of expected width of the $95\%$ Wald confidence interval for $n = 50$ uniformly over $\pi \in (0,1)$ (blue bottom axis with blue area) and variance-stabilized over $\phi = \arcsin(\sqrt{\pi})\in (0, \pi/2)$ (red top axis with red area).}
\label{integral}
\end{figure}

Figure~\ref{integral} illustrates a summary measure that integrates uniformly over $\pi$ and one that integrates on the variance-stabilized scale of $\pi$. This idea was motivated by the poor performance of some CIs for extreme true proportions near $0$ or $1$. By transforming the proportions with the variance-stabilizing transformation
\begin{equation*}
\phi = \arcsin(\sqrt{\pi})
\end{equation*}
and integrating on that scale, more weight is given to these extreme cases. More precisely, instead of integrating a function $f(\pi)$ from $0$ to $1$, the function $f\{\sin^2(\phi)\}$ is integrated from $0$ to $\pi/2$ (where $\pi$ denotes the mathematical constant 3.14159...) as
\begin{equation*}
f(\pi) = f\{\sin^2(\arcsin(\sqrt{\pi}))\} = f\{\sin^2(\phi)\}.
\end{equation*}
Figure~\ref{integral} illustrates that the integral on the variance-stabilized scale attributes more weight to the boundaries. It is illustrated exemplary for the expected width of a $95\%$ Wald CI with $n=50$. Note that the $x$-axis for the transformed integral is shifted to the left. The weight density of the variance-stabilizing transformation is the first derivative which is the density of the Jeffreys prior $\Be(1/2,1/2)$ \citep[p.~187]{HeldSabanesBove}. Consequently, the integral on the variance-stabilized scale is the integrated risk for the Jeffreys prior (up to the constant factor $\pi/2$). The integral on the uniform scale is the integrated risk for the uniform prior.

\subsection{Asymptotical expected interval score}

The (actual) coverage probability is asymptotically equal to the nominal coverage probability $\gamma$. Similarly, an asymptotical reference value can be derived for the expected interval score. It is useful for better visualizations in Section~\ref{sec:res}.

The normal approximation of the binomial distribution $X\sim \Bin(n, \pi)$ by the central limit theorem, $X/n\sima \N(\pi, \pi(1-\pi)/n)$, can be used to construct a $\gamma\cdot 100\%$ CI with limits
\begin{equation*}
L = \pihat - q_\alpha \cdot \sigma \quad \text{and} \quad U = \pihat + q_\alpha \cdot \sigma
\end{equation*}
under asymptotical normality of $\pihat = X/n$ with $\sigma = \sqrt{\pi(1-\pi)/n}$ the known standard deviation and $q_\alpha$ the $1-\alpha/2$ quantile of $\N(0,1)$. The interval score of this CI is
\begin{equation*}
\IS_\alpha(L, U, \pi) = 2 q_\alpha \sigma + \frac{2}{\alpha}(L - \pi)\ind(L>\pi) + 
\frac{2}{\alpha}(\pi - U)\ind(U<\pi)
\end{equation*}
and by linearity, the expected interval score is
\begin{equation*}
\Exp\{\IS_\alpha(L, U, \pi)\} = 2 q_\alpha \sigma + \frac{2}{\alpha}\Exp\{(L - \pi)\ind(L-\pi>0)\} + 
\frac{2}{\alpha}\Exp\{(\pi - U)\ind(\pi-U>0)\}.
\end{equation*}
Under the asymptotical normal distribution of $X/n$, both $L-\pi$ and $\pi-U$ have a normal distribution $\N(-q_\alpha \sigma, \sigma^2)$. Defining $Y \sim \N(-q_\alpha \sigma, \sigma^2)$, the asymptotical value of the expected interval score is
\begin{align*}
\Exp\{\IS_\alpha(L, U, \pi)\} 
&= 2 q_\alpha \sigma + 2\frac{2}{\alpha}\Exp\{Y\ind(Y>0)\}\\
&= 2 q_\alpha \sigma + 2\frac{2}{\alpha}\Exp\{Y \given Y>0\} \Prob(Y>0)\\
&= 2 q_\alpha \sigma + 2\Exp\{Y \given Y>0\}\\
&= 2 q_\alpha \sigma + 2\left(-q_\alpha \sigma + \frac{\sigma\phi(q_\alpha)}{1-\Phi(q_\alpha)}\right)\\
&= \frac{2\sigma\phi(q_\alpha)}{1-\Phi(q_\alpha)},
\end{align*}
where $\phi$ and $\Phi$ are the density and distribution functions of a standard normal. The derivation uses conditional expectations and that $\Prob(Y>0)=\alpha/2$. The expectation of a truncated normal random variable $Y \given Y>0$ is taken from \cite{JohnsonKotz}. Likewise, the asymptotical value of the expected width is
\begin{equation*}
\Exp\{\W(L, U)\} = 2 q_\alpha \sigma.
\end{equation*}

\begin{figure}[h]
\begin{knitrout}
\definecolor{shadecolor}{rgb}{0.969, 0.969, 0.969}\color{fgcolor}

{\centering \includegraphics[width=\maxwidth]{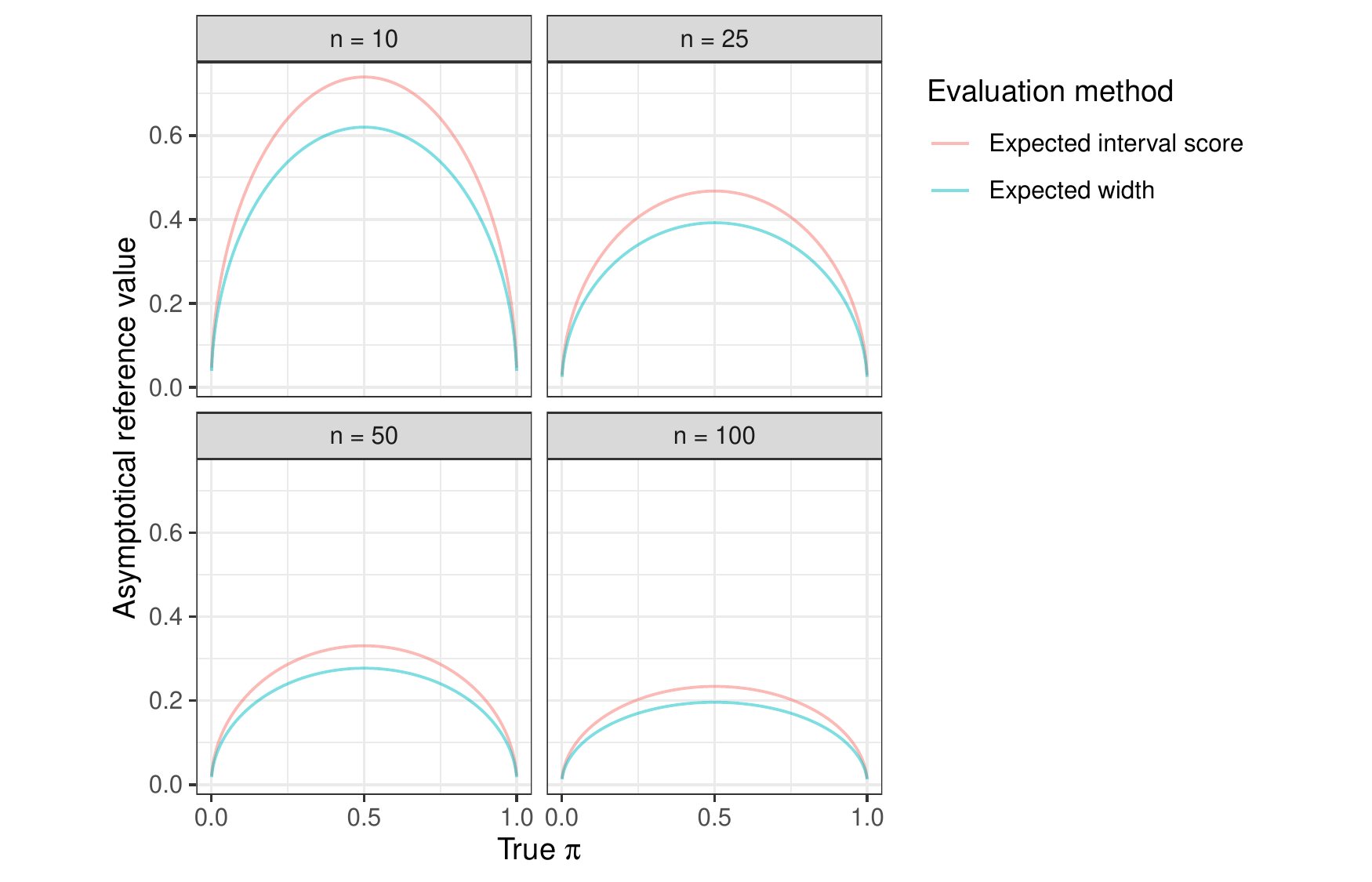} 

}

\end{knitrout}
\caption{Asymptotical expected interval score and asymptotical expected width of a confidence interval under asymptotical normality of a binomial sample as a function of $\pi$ for $\gamma = 0.95$ and $n\in \{10, 25, 50, 100\}$.}
\label{asymptotics}
\end{figure}

The asymptotical expected interval score and expected width are concave curves as a function of the true proportion $\pi$, as can be seen in Figure~\ref{asymptotics} for $\gamma=0.95$ and different values of $n$. The expected interval score (not asymptotical) has the same curve shape, similar to the expected widths in Figure~\ref{covwidth}. Because of this curvature, the differences between the CIs are hardly visible. By taking the difference to the asymptotical reference curve, the curvature can be removed and differences between the CIs are easier to visualize.

\section{Results} \label{sec:res}

\begin{figure}
\begin{knitrout}
\definecolor{shadecolor}{rgb}{0.969, 0.969, 0.969}\color{fgcolor}

{\centering \includegraphics[width=\maxwidth]{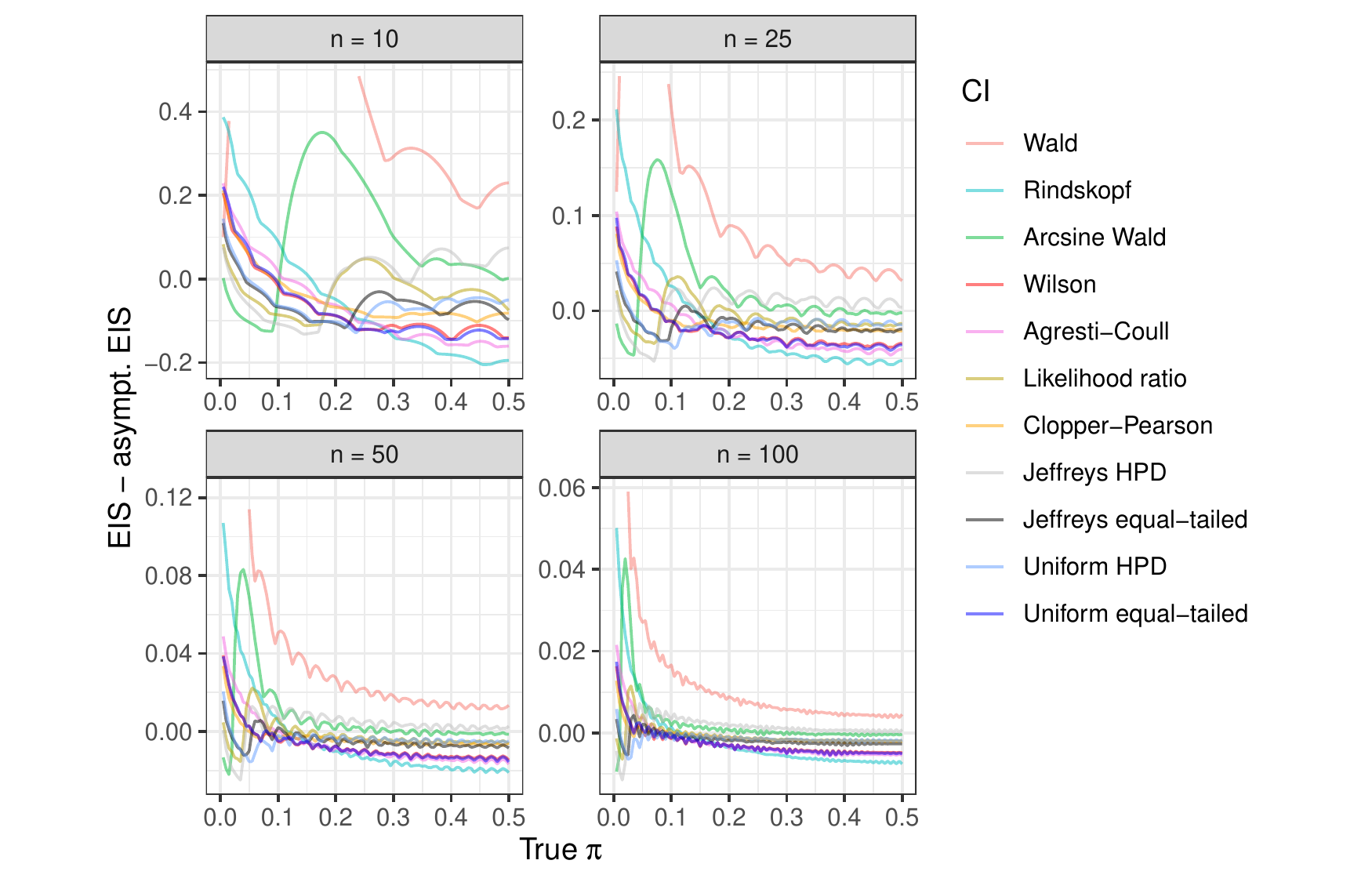} 

}

\end{knitrout}
\caption{Difference between expected interval score (EIS) and asymptotical EIS for each confidence interval (CI) method as a function of $\pi$ for $\gamma = 0.95$ and $n\in \{10, 25, 50, 100\}$.}
\label{IS_n}
\end{figure}

Figure~\ref{IS_n} compares the expected interval scores of $95\%$ CIs of binomial samples for different sample sizes $n$ as a function of the true proportion $\pi$ (symmetric around $\pi=0.5$). What is shown is the difference between the expected interval score and the asymptotical reference value of the expected interval score (under normality). Lower scores are better, so the Wald CI is the worst CI in terms of the expected interval score. The scores would be even larger but to see the differences between the CIs better, these parts have been cut off in the plots. The arcsine Wald CI performs similarly bad except near the boundaries $0$ and $1$ of $\pi$ where it even has the best score. It is the other way around for the Rindskopf and the Agresti-Coull CIs which are the best for $\pi=0.5$ but the worst or second worst for $\pi$ near the boundaries. Similarly, the HPD intervals are better than the equal-tailed intervals for $\pi=0.5$ but worse at the boundaries. The Wilson and the uniform equal-tailed CIs are very similiar for all $n$. The likelihood ratio CI is worse than these two except at the boundaries.

\begin{figure}
\begin{knitrout}
\definecolor{shadecolor}{rgb}{0.969, 0.969, 0.969}\color{fgcolor}

{\centering \includegraphics[width=\maxwidth]{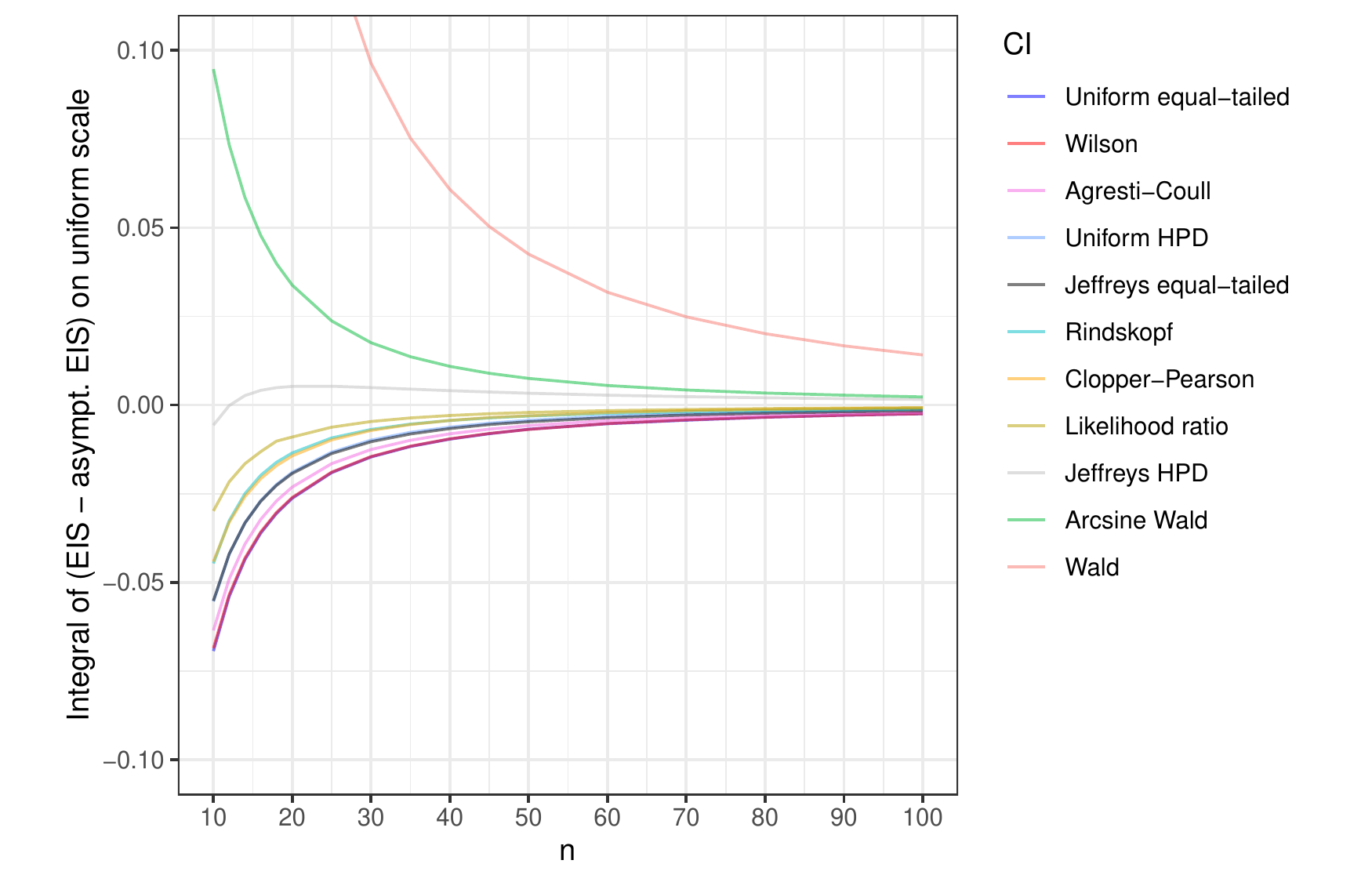} 

}

{\centering \includegraphics[width=\maxwidth]{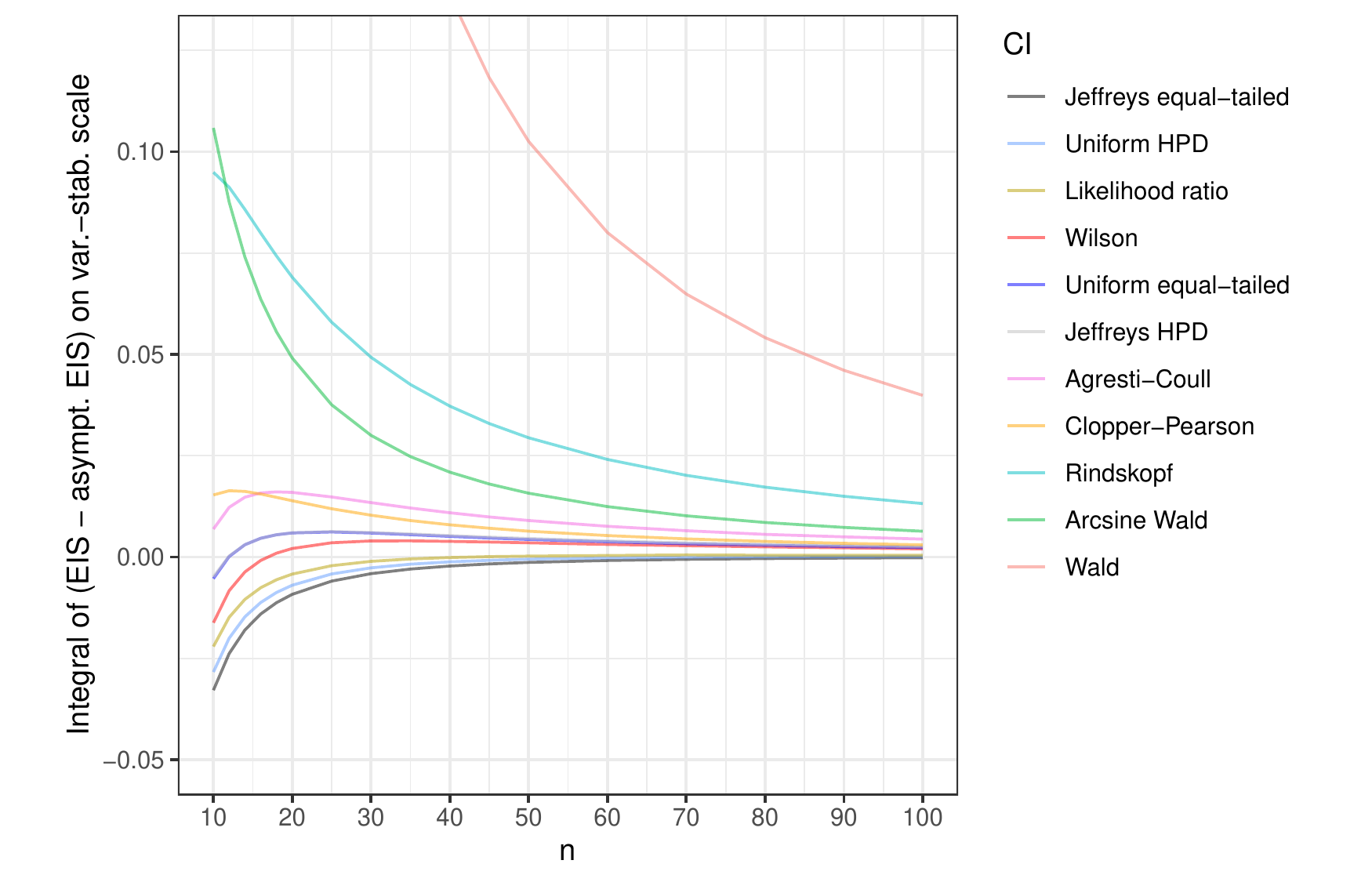} 

}

\end{knitrout}
\caption{Uniform (top) and variance-stabilized (bottom) integral of difference between expected interval score (EIS) and asymptotical EIS for each confidence interval (CI) method as a function of $n$ for $\gamma = 0.95$. The legends are ranked for $n=10$.}
\label{int}
\end{figure}

The uniform integral over all possible true proportions $\pi\in (0,1)$ is used as a summary measure. Also here, for visualization reasons, the function that is integrated is the difference between the expected interval score and the asymptotical reference of the expected interval score. With this summary measure, a clear ranking is obtained, meaning that one number can be compared across CIs for each sample size $n$ and confidence level $\gamma$. The top plot in Figure~\ref{int} compares the uniform integrals of $95\%$ CIs as a function of $n$. The legend is ranked for $n=10$ where the best CI is at the top. The only difference in the ranking for $n=100$ is that the HPD interval changes to the sixth place. The three best CIs are the uniform equal-tailed, the Wilson and the Agresti-Coull (in decreasing order). The three worst CIs are the Jeffreys HPD, the arcsine Wald and the Wald (in decreasing order). All CIs converge to the same value for increasing $n$ but not all at the same speed.

The bottom plot in Figure~\ref{int} is a comparison of the CIs by the variance-stabilized integral. The ranking of the CIs does indeed change compared to the uniform integral. Now, for $n=10$, the Jeffreys equal-tailed, the uniform HPD and the likelihood ratio CIs are at the top of the list (in decreasing order), followed by the Wilson CI. The arcsine Wald and Wald CIs are still at the bottom of the list. For small $n$, the ranking changes between the Rindskopf and the arcsine Wald CIs and between the Agresti-Coull and the Clopper-Pearson CIs. The good performance of the Jeffreys equal-tailed CI can be explained by the connection between the Jeffreys prior and the variance-stabilizing transformation (see Subsection~\ref{subsec:integral} and Appendix~\ref{sec:intervals}).

\section{Discussion} \label{sec:discussion}

The interval score is a new criterion to evaluate CIs. In a frequentist setting, the expected interval score can be compared across different types of CIs. It is intuitively appealing since it combines the width as a measure of sharpness and the coverage as a measure of over- and underestimation. One could even look at over- and underestimation separately if this would be of interest \citep{GneitingBracher}. What is important is that non-coverage is penalized more with increasing minimal distance of the true proportion to the CI and with larger confidence levels. So, the interval score does not only address non-coverage per se but the \emph{amount} of non-coverage. It is in particular this new idea that makes the interval score a proper scoring rule for central prediction intervals and consequently solves a paradox of another score that tried to combine width and coverage \citep{GneitingRaftery}. Comparison of CIs using the expected interval score is a new application of the interval score, already proposed in \cite{GneitingRaftery} but to the best of our knowledge not yet applied.

The integral summary measure is a way of averaging the considered measure, in our case the expected interval score, over different settings of true proportions $\pi$ (but for fixed sample size and confidence level). This concept is closely related to integrated risk in Bayesian decision theory. Similarly, the mean coverage probability is related to the uniform integral because usually it assumes an underlying uniform distribution of $\pi$ \citep{Newcombe2013}. The advantage of the integral summary measure is that one can integrate on different scales. By integrating on the variance-stabilized scale, more weight can be given to extreme cases. Poor performance in extreme cases is a problem for many CIs for the binomial proportion. It should be taken into account since according to \citet[p.~17]{Tuyl}, ``the most important property of a method is that it produces sensible intervals for any possible data outcome'', and according to \citet[p.~178]{Jaynes}, ``the merits of any statistical method are determined by the results it gives when applied to specific problems'' (quoted in \cite{Newcombe2013}). Indeed, in practice, low or zero counts can happen if the sample size is small or for rare diseases. Large counts nearly equal to $n$ can happen \eg when estimating the sensitivity or the specificity of a diagnostic test. Also in these cases, a CI method should produce a sensible interval. The variance-stabilized integral summary measure is a new idea that results in different recommendations since usually different CIs perform well in extreme cases and non-extreme cases.

We have derived a new asymptotical result for the expected interval score in the binomial setting. 
Subtracting the same asymptotical reference value from the expected interval score of all CI methods under consideration does not change their ranking. This approach facilitates the visibility of differences in graphs, especially for small $n$. In passing we note that asymptotical results can also be derived for the variance of the interval score, but this turned out to be less useful for the comparison of CIs.  


These new techniques to compare CIs for the binomial proportion result in two clear rankings. Based on the uniform integral summary measure of the expected interval score, we recommend the Wilson and the uniform equal-tailed as the CIs with the best performance. If we want to give more weight to performance in extreme cases, the Jeffreys equal-tailed or uniform HPD CIs should be chosen based on the variance-stabilized integral summary measure. Generally, the Wald CI is the worst of the compared methods. Its good performance in terms of width for $\pi$ near the boundaries influences the interval score less than the bad coverage. Also in the literature there is consensus that the Wald CI should generally not be used \citep{Brown2001,Gillibert}. With a separate evaluation of coverage probabilities and expected widths, \citet{Brown2001} has also recommended the Wilson and Jeffreys equal-tailed CIs. The Agresti-Coull CI has sometimes been preferred because it is simpler and performs equally well for larger $n$ \citep{Brown2001,PiresAmado}. The mid-P Clopper-Pearson CI, which is approximately equal to the Jeffreys equal-tailed CI \citep[p.~114]{Brown2001}, has been recommended because of its good performance and the balance of one-sided errors \citep{Newcombe2013,Gillibert}. Also \citet{Newcombe1998} has recommended the Wilson or mid-P Clopper-Pearson CIs if a nominal mean coverage probability is desired.

It is remarkable that the Bayesian equal-tailed CI with either a uniform prior for uniform weighting or Jeffreys' prior for ``Jeffreys'' weighting (since the density of the Jeffreys prior is the weight density of the variance-stabilizing transformation) performs best. The close correspondence of the prior and the weight function suggests that the 
equal-tailed Bayesian CI with the corresponding prior will have good properties also for other weight function. If we aim for a CI that performs well for both weightings, then we would favor the Wilson and uniform HPD CIs because they are under the top four in both scenarios. 
Another desired property of a CI is invariance under general one-to-one parameter transformations. Such CIs are the Wilson as a score CI, the likelihood ratio CI and Bayesian intervals with Jeffreys' prior \citep{HeldSabanesBove}. Among those the Wilson CI performs particularly well under both weightings. 

The major advantage of the interval score as an evaluation method is that it is a proper scoring rule. According to \citet{GneitingRaftery}, it solves that ``the question of measuring optimality (either frequentist or Bayesian) of a set estimator against a loss criterion combining size and coverage does not yet have a satisfactory answer'', pointed out by \citet[p.~141]{CasellaHwangRobert}. However, it is strictly speaking a proper scoring rule in a different setting, namely probabilistic interval forecasts. To define (and prove) propriety in the setting of frequentist interval estimates is in our view still an open question. A solution could be to think in the direction of \emph{confidence distributions} described \eg in \cite{XieSingh}. A second issue is that the interval score is only a proper scoring rule for \emph{central} prediction intervals and consequently CIs. In our comparison, only the Bayesian equal-tailed CIs are central whereas all other CIs are only asymptotically central. Since all CIs have good properties asymptotically and a comparison is interesting rather for smaller $n$ that are used in practice, this is a limitation.

\begin{figure}
\begin{knitrout}
\definecolor{shadecolor}{rgb}{0.969, 0.969, 0.969}\color{fgcolor}

{\centering \includegraphics[width=\maxwidth]{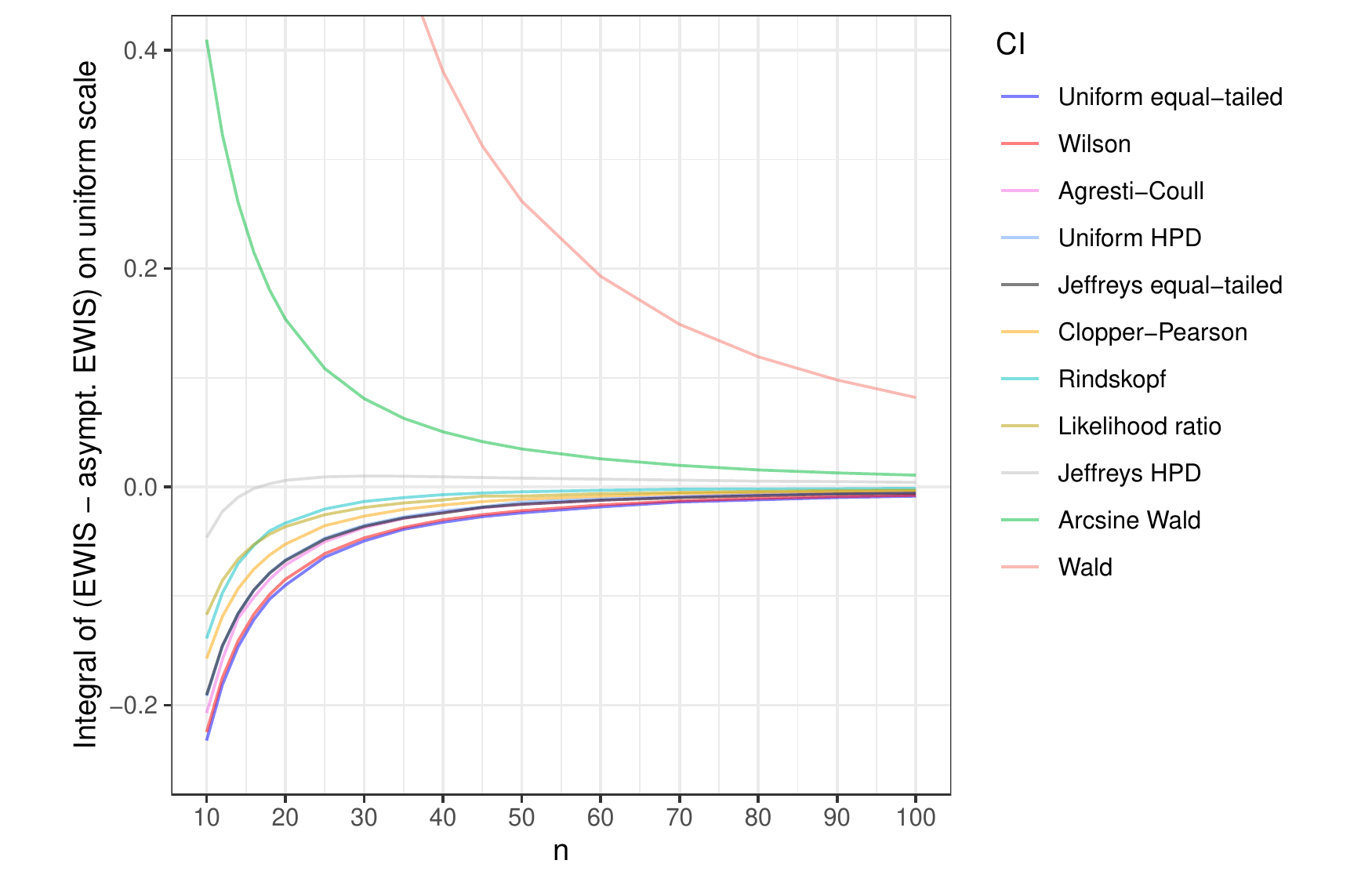} 

}

{\centering \includegraphics[width=\maxwidth]{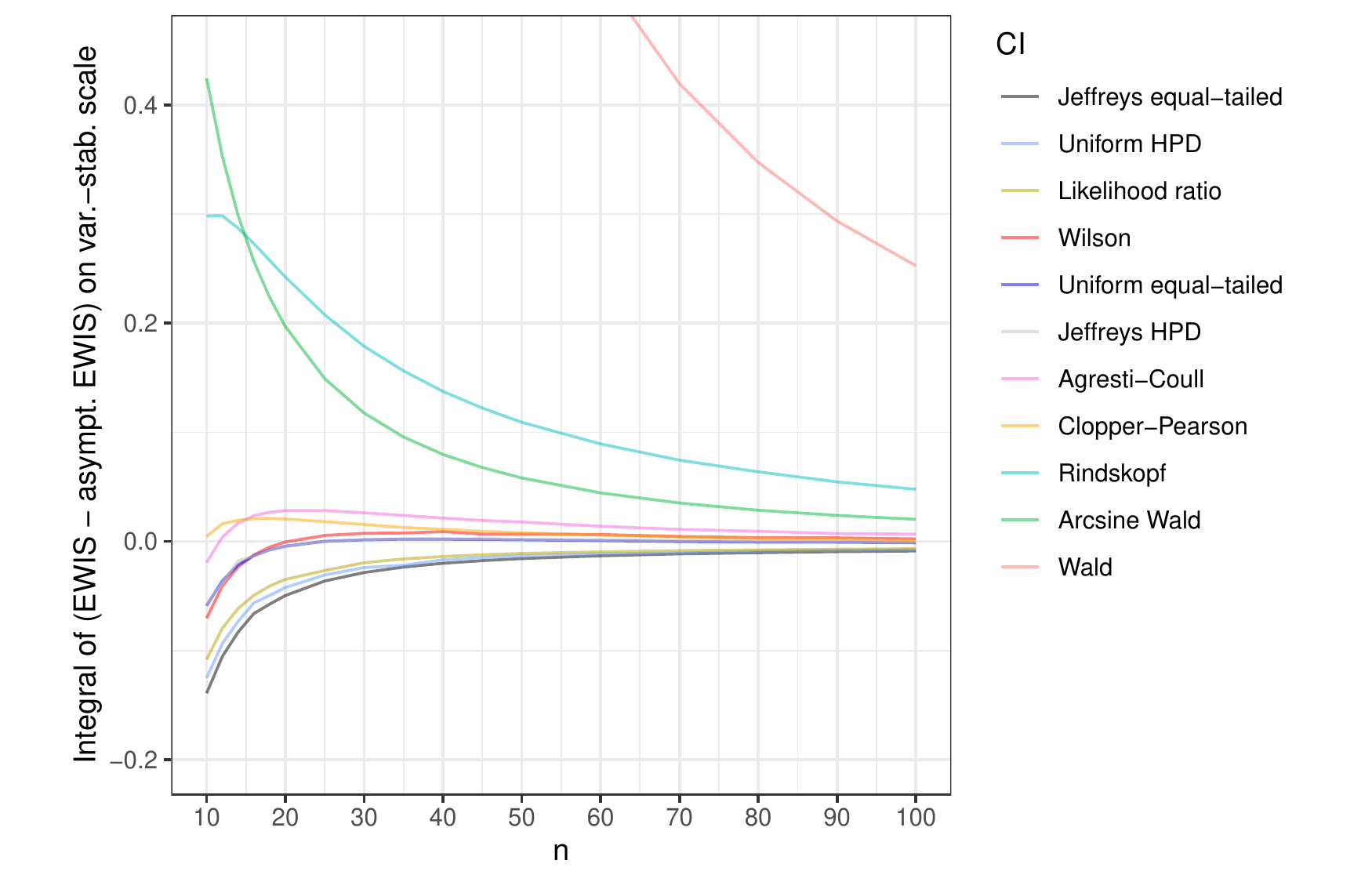} 

}

\end{knitrout}
\caption{Uniform (top) and variance-stabilized (bottom) integral of difference between expected weighted interval score (EWIS) and asymptotical EWIS for each confidence interval (CI) method as a function of $n$ combining confidence levels $0.9$, $0.95$ and $0.99$ for weights $1$. The legends are ranked for $n=10$.}
\label{int_WIS}
\end{figure}

Our recommendations are based on integral rankings for a fixed sample size and confidence level. For sample sizes $n$ between $10$ and $100$, the rankings of $95\%$ CIs do virtually not depend on $n$. The confidence level, however, can have a larger impact \citep{Thulin}, although most comparisons only consider $95\%$ CIs for the binomial proportion since these are the standard in practice. A way to average over different confidence level settings is the \emph{weighted interval score} introduced in \cite{GneitingBracher}. It combines multiple confidence levels in a weighted sum of the interval scores for these levels. We combined the three levels $0.9$, $0.95$ and $0.99$ with weights equal to $1$. Due to linearity, the expected weighted interval score is the weighted sum of the expected interval scores. The only differences of the integral rankings in Figure~\ref{int_WIS} compared to the $0.95$ case in Figure~\ref{int} are: Based on the uniform integral, the Rindskopf CI worsens towards the likelihood ratio CI and the Agresti-Coull CI worsens towards the Jeffreys equal-tailed CI. Based on the variance-stabilized integral, the Wilson CI is worse than the uniform equal-tailed CI for $n > 15$.

One possible extension of our results would be to compare a certain version of the weighted interval score that approximates the continuous ranked probability score (CRPS) as described in \cite{GneitingBracher}. The CRPS version is particularly suitable for Bayesian intervals because of similarities to a probabilistic forecasting setting. Finally, the new approach of using the expected interval score to compare CIs could also be applied to other parameters than the binomial proportion. For example to heterogeneity variances in meta-analysis, where it could be valuable to better understand the different CIs.

\section*{Software} \label{sec:software}

We implemented all methods, including confidence and credible intervals, in the \R language for statistical computing \citep[Version 4.1.2]{R} and the code to reproduce the results is available at \url{https://github.com/lisajohanna/intervalscoreCode}. Parts of this implementation are based on code from \cite{HeldSabanesBove} available at \url{https://github.com/lheld/HSB} and we used the \texttt{ggplot2}, \texttt{scales} and \texttt{ggpubr} packages to create graphics \citep{ggplot2,scales,ggpubr}. The sample size $n$ is small enough such that no simulations are necessary and the measures can be calculated analytically. Since the integrals rely on numerical integration, we used parallel computing with the \texttt{foreach} and \texttt{doParallel} packages to speed up the computations \citep{doParallel,foreach}.

\bibliographystyle{mywiley_unsrt}
\bibliography{biblio}

\appendix

\section{Computation of confidence and credible intervals} \label{sec:intervals}

All considered confidence intervals except the Clopper-Pearson are constructed from approximate pivots:
\begingroup
\allowdisplaybreaks
\begin{align*}
\text{Wald statistic}\quad &\frac{\frac{X}{n}-\pi}{\sqrt{\frac{X/n(1-X/n)}{n}}} \sima \N(0,1),\\
\text{score statistic}\quad &\frac{\frac{X}{n}-\pi}{\sqrt{\frac{\pi(1-\pi)}{n}}} \sima \N(0,1),\\
\text{likelihood ratio statistic}\quad &-2\log\left\{\frac{L(\pi)}{L(\pihat)}\right\} \sima \chi^2(1),
\end{align*}
where $\N(0,1)$ denotes the standard normal distribution, $\chi^2(1)$ denotes the chi-squared distribution with one degree of freedom and the symbol $\sima$ means ``is asymptotically distributed as'' in the sense of convergence in distribution for $n\to \infty$. These statistics use quantities from likelihood inference:
\allowdisplaybreaks
\begin{align*}
\text{likelihood function}\quad &L(\pi) = {n\choose x} \pi^{x}(1-\pi)^{n-x},\\
\text{maximum likelihood estimate (MLE)}\quad &\pihat = \frac{x}{n},\\
\text{standard error}\quad &\se(\pihat) = \sqrt{\frac{\pihat(1-\pihat)}{n}}.
\end{align*}
\endgroup
We denote by $q_\alpha$ the $1-\alpha/2$ quantile of $\N(0,1)$ where $\alpha=1-\gamma$.

For the credible intervals, we consider two beta priors. Beta priors have the appropriate support $(0,1)$ and are conjugate for the binomial proportion, as the $\Be(\alpha,\beta)$ prior leads to the $\Be(\alpha + x, \beta + n-x)$ posterior.
\begin{enumerate}
\item Using the \emph{uniform} prior $\pi\sim \Unif(0,1) = \Be(1,1)$, the posterior distribution is
\begin{equation*}
\pi \given x \sim \Be(1 + x, 1 + n-x).
\end{equation*}
This prior is non-informative (on the scale of $\pi$) since any value between $0$ and $1$ is equally likely. However, it would not be uniform anymore for nonlinear transformations of $\pi$.
\item Using the \emph{Jeffreys} prior $\pi \sim \Be(1/2,1/2)$, the posterior distribution is
\begin{equation*}
\pi \given x \sim \Be(1/2 + x, 1/2 + n-x).
\end{equation*}
The Jeffreys prior is defined as the prior that is proportional to $\sqrt{J(\pi)}$, where $J(\pi)$ is the expected Fisher information, and it is invariant under reparametrization. That means the prior of a transformed parameter is still a Jeffreys prior. \citet[p.~186--187]{HeldSabanesBove} provide an argument why this is a non-informative prior. 
\end{enumerate}

\subsection{Clopper-Pearson}
The (discrete) realizations $x$ will lie between some $x_1$ and $x_2$ with a probability of \emph{at least} $\gamma$. The $\gamma\cdot 100\%$ Clopper-Pearson confidence interval \citep{ClopperPearson} inverts the determining inequalities for this interval for $x$ to obtain an interval for the (continuous) parameter $\pi$. The limits $L$ and $U$ are derived from the two equations \citep{PiresAmado}
\begin{equation}\label{eq:exactbinom}
\sum_{j=x}^{n}{n \choose j}L^{j}(1-L)^{n-j}=\frac{\alpha}{2} \quad \text{and} \quad 
\sum_{j=0}^{x}{n \choose j}U^{j}(1-U)^{n-j}=\frac{\alpha}{2}.
\end{equation}
The quantities in \eqref{eq:exactbinom} are interpreted as $\Prob(X\geq x)$ for $X\sim \Bin(n,L)$ and $\Prob(X\leq x)$ for $X\sim \Bin(n,U)$. The relation
\begin{equation*}
\sum_{j=x}^{n}{n \choose j}\pi^{j}(1-\pi)^{n-j}=\int_0^\pi f_b(t) dt
\end{equation*}
to the beta density function $f_b$ of $\Be(x, n-x+1)$ is used to solve \eqref{eq:exactbinom} for the limits
\begin{equation}\label{eq:CPlimits} 
L = b_{(1-\gamma)/2}(x, n-x+1) \quad \text{and} \quad U = b_{(1+\gamma)/2}(x+1, n-x) \quad
\text{for} \quad 0<x<n,
\end{equation}
where $b_\gamma(\alpha, \beta)$ is the $\gamma$ quantile of $\Be(\alpha, \beta)$. The limits \eqref{eq:CPlimits} can be interpreted in a Bayesian way with two different priors. They are equal to the lower, respectively upper, limit of an equal-tailed interval with an improper $\Be(0,1)$, respectively $\Be(1,0)$, prior.

For $x=0$ and $x=n$, the lower, respectively upper limit in \eqref{eq:CPlimits} is improper (one parameter is $0$). In these cases, the solutions are calculated directly:
\begin{align*}
L = 0 \quad \text{and} \quad U = 1-(\alpha/2)^{1/n} \quad &\text{for} \quad x = 0,\\
L = (\alpha/2)^{1/n} \quad \text{and} \quad U = 1 \quad &\text{for} \quad x = n.
\end{align*}

The Clopper-Pearson interval is known as an ``exact'' interval because \eqref{eq:exactbinom} uses the exact distribution $X\sim \Bin(n, \pi)$. However, it does not have exact coverage probability equal to $\gamma$. On the contrary, the minimum coverage probability (for any true proportion $\pi$) is \emph{at least} $\gamma$. Hence, it is conservative. 

\subsection{Wilson}
Based on the standard normal approximation of the score statistic, the $\gamma\cdot 100\%$ Wilson confidence interval \citep{Wilson} is the set of all parameter values $\pi$ that satisfy
\begin{equation*}
\pi^2\left(n^2+nq_\alpha^2\right)+\pi\left(-2nx-nq_\alpha^2\right)+x^2=0.
\end{equation*}
Solving this quadratic equation yields the limits
\begin{equation*}
\frac{x + q_\alpha^2/2}{n + q_\alpha^2} \pm \frac{q_\alpha \sqrt{n}}{n + q_\alpha^2} \sqrt{\pihat(1-\pihat) + \frac{q_\alpha^2}{4n}}.
\end{equation*}
The midpoint of the Wilson interval is the relative proportion of successes after adding $q_\alpha^2/2$ successes and non-successes to the sample, called \emph{shrinkage estimator} \citep{Newcombe2013}.

\subsection{Wald}
Based on the standard normal approximation of the Wald statistic, the limits of the $\gamma\cdot 100\%$ Wald confidence interval \citep{Wald} have the simple form
\begin{equation*}
\pihat \pm q_\alpha\cdot \se(\pihat) \quad \text{with} \quad 
\se(\pihat) = \sqrt{\frac{\pihat(1-\pihat)}{n}}.
\end{equation*}
They may fall outside the range $(0,1)$ for $\pi$. This problem is referred to as \emph{overshoot} or \emph{boundary violation} and happens for small or large $x$. For $95\%$ confidence, overshoot occurs whenever $x=1$ or $x=2$, and also when $x=3$ except when $n<14$ \citep{Newcombe1998}. It happens more often for large confidence levels due to the wider confidence interval. Overshoot is truncated to $(0,1)$, as it is usually done in the literature. Truncation cannot affect coverage properties but limits $0$ or $1$ are unsatisfactory since they are uninterpretable if $0<x<n$ \citep{Newcombe2013}.

Since the standard error is $0$ for the extreme cases $x=0$ ($\pihat=0$) and $x=n$ ($\pihat=1$), the Wald interval is a \emph{degenerate} or \emph{zero width interval} in these cases (for any confidence level).

\subsection{Rindskopf}
A $\gamma\cdot 100\%$ Wald confidence interval is calculated for the logit transformed parameter
\begin{equation*}
\phi = \logit(\pi) = \log\left(\frac{\pi}{1-\pi}\right)
\end{equation*}
with
\begin{equation*}
\phihat = \log\left(\frac{x + 0.5}{n-x + 0.5}\right) \quad \text{and} \quad 
\se(\phihat) = \sqrt{\frac{1}{x + 0.5}+\frac{1}{n-x + 0.5}}.
\end{equation*}
The adjustment of adding $0.5$ successes and non-successes ensures that also for the cases $x=0$ and $x=n$ (with otherwise infinite MLE and standard error) an interval can be computed. Since the scale of $\phi$ is $(-\infty, \infty)$, this interval is boundary respecting. Back-transformation to the scale of $\pi$ with the inverse logit function
\begin{equation*}
\pi = \expit(\phi) = \frac{\exp(\phi)}{1+\exp(\phi)}
\end{equation*}
yields the Rindskopf confidence interval. We call it \emph{Rindskopf} because the used adjustment was suggested by \citet{Rindskopf}.

\subsection{Arcsine Wald}
A $\gamma\cdot 100\%$ Wald confidence interval is calculated for the variance-stabilizing transformation
\begin{equation*}
\phi = \arcsin\left(\sqrt{\pi}\right)
\end{equation*}
with
\begin{equation*}
\phihat = \arcsin\left(\sqrt{\pihat}\right) \quad \text{and} \quad 
\se(\phihat) \approx \frac{1}{\sqrt{4n}}.
\end{equation*}
It is called variance-stabilizing since the variance of $\phihat$ is asymptotically independent of the parameter $\phi$. Since the scale of $\phi$ is $(0, \pi/2)$, where $\pi$ for once means the mathematical constant, this interval may overshoot (\eg for $x=0$ and $x=n$). In these cases, the interval is truncated to $(0, \pi/2)$. Back-transformation to the scale of $\pi$ with the inverse function
\begin{equation*}
\pi = \sin^2(\phi)
\end{equation*}
yields the arcsine Wald confidence interval.

\subsection{Agresti-Coull}
The limits of the $\gamma\cdot 100\%$ Agresti-Coull confidence interval are
\begin{equation*}
\tilde{\pi} \pm q_\alpha \cdot \sqrt{\frac{\tilde{\pi}(1-\tilde{\pi})}{n+4}} \quad \text{with} \quad 
\tilde{\pi}=\frac{x+2}{n+4}.
\end{equation*}
It was called the ``add two successes and two failures'' adjusted Wald confidence interval by \citet{AgrestiCoull}. It was motivated by the finding that for $\gamma=0.95$, where $q_\alpha^2\approx 4$, the midpoint $\tilde{\pi}$ is approximately equal to the midpoint of the Wilson interval. Although this is only true for $\gamma=0.95$, the same adjustment is used for any confidence level. The midpoint $\tilde{\pi}$ is identical to the Bayes estimate (mean of the posterior distribution) for a $\Be(2,2)$ prior.

\subsection{Likelihood ratio}
The $\gamma\cdot 100\%$ likelihood ratio confidence interval uses the right tail of the $\chi^2(1)$ approximation of the likelihood ratio statistic to derive the condition
\begin{equation}\label{lr_cond}
-2\log\left\{\frac{L(\pi)}{L(\pihat)}\right\} \leq \chi^2_{\gamma}(1),
\end{equation}
where $\chi^2_{\gamma}(1)$ is the $\gamma$ quantile of the $\chi^2(1)$ distribution. It consists of all parameter values $\pi$ that satisfy~\eqref{lr_cond}. The limits are calculated numerically using \R function \texttt{uniroot}. Only one solution is obtained for the cases $x=0$ and $x=n$ where the second limit is set to $0$, respectively $1$.

\subsection{Equal-tailed}
The limits of the $\gamma\cdot 100\%$ equal-tailed credible interval are the $(1-\gamma)/2$ and $(1+\gamma)/2$ quantiles of the posterior distribution. A probability mass of $\alpha/2$ is cut off from both tails of the posterior distribution. For the two priors, the limits are:
\begin{enumerate}
\item Uniform prior:
$\quad b_{(1-\gamma)/2}(1 + x, 1 + n-x) \quad \text{and} \quad b_{(1+\gamma)/2}(1 + x, 1 + n-x).$
\item Jeffreys prior:
$\quad b_{(1-\gamma)/2}(1/2 + x, 1/2 + n-x) \quad \text{and} \quad b_{(1+\gamma)/2}(1/2 + x, 1/2 + n-x).$
\end{enumerate}

\subsection{Highest posterior density}
The highest posterior density (HPD) interval is the unique (for the chosen prior) $\gamma\cdot 100\%$ credible interval $[l,u]$ that fulfills the condition
\begin{equation*}
f(\pi \given x)\geq f(\tilde{\pi} \given x)
\end{equation*}
for all $\pi\in [l,u]$ and all $\tilde{\pi}\notin [l,u]$. It consists of all the parameter values with the highest posterior density until they reach a probability mass of $\gamma$, making it the smallest interval that has mean coverage probability $\gamma$ under the specified prior. The limits are calculated numerically except for the two extreme cases. Since the posterior density is monotone decreasing for $x=0$, the lower limit is $0$ and the upper limit is the $\gamma$ quantile of the posterior distribution. Since the posterior density is monotone increasing for $x=n$, the upper limit is $1$ and the lower limit is the $1-\gamma$ quantile of the posterior distribution.

\end{document}